\documentclass[conference]{IEEEtran}
\IEEEoverridecommandlockouts
\usepackage{cite}
\usepackage{amsmath,amssymb,amsfonts}
\usepackage{algorithmic}
\usepackage{graphicx}
\usepackage{textcomp}
\usepackage{xcolor}
\def\BibTeX{{\rm B\kern-.05em{\sc i\kern-.025em b}\kern-.08em
    T\kern-.1667em\lower.7ex\hbox{E}\kern-.125emX}}
\begin{document}

\title{Machine Learning Fairness in Justice Systems: Base Rates, False Positives, and False Negatives\\
}

\author{
\IEEEauthorblockN{Jesse Russell}
\IEEEauthorblockA{
Reno, NV, USA \\
jesse.r.russell@gmail.com, https://orcid.org/0000-0002-4856-2462}
}

\maketitle

\begin{abstract}
Machine learning best practice statements have proliferated, but there is a lack of consensus on what the standards should be. For fairness standards in particular, there is little guidance on how fairness might be achieved in practice. Specifically, fairness in errors (both false negatives and false positives) can pose a problem of how to set weights, how to make unavoidable tradeoffs, and how to judge models that present different kinds of errors across racial groups. This paper considers the consequences of having higher rates of false positives for one racial group and higher rates of false negatives for another racial group. The paper examines how different errors in justice settings can present problems for machine learning applications, the limits of computation for resolving tradeoffs, and how solutions might have to be crafted through courageous conversations with leadership, line workers, stakeholders, and impacted communities.
\end{abstract}


\section{Introduction}
As machine learning best practice and ethical practice guidelines have proliferated, there has been a lack of consensus of what should be included in the rules. Each set of ethical guidelines is unique. Even so, there have been some common themes. One recurrent axiom is that machine learning models should be fair and equitable. Fairness, though, is not always clear and there is no consensus definition. Many papers have catalogued different definitions of fairness that might be applied to machine learning. For example, Verma and Rubin \cite{b1} identify more than twenty different forms of equity that have been proposed for machine learning.
Fairness and equity in machine learning are especially critical when applied to justice settings. Machine learning that is used to create decision support tools around which people should penetrate the system further, or should be diverted, or should receive a particular disposition, or should be placed in a secured facility should be fair. These kinds of decisions, with all their consequences, must be fair to fulfil the idea of justice. It is a standard set out in the U.S. Constitution. The Equal Protection Clause of the Fourteenth Amendment establishes that government entities, including justice systems, must treat people in the same situation in the same way. While the ethical standard is clear in this regard, it is not clear how machine learning can uphold this Constitutional right.
Most ethical machine learning guidelines provide a stance on fairness without granular guidance on how to work in a setting that is not itself fair or with data that are tainted with prejudice bias from history or from structural inequality. Guidance from IBM, for example, says that, "AI must be designed to minimize bias and promote inclusive representation" \cite{b2}. There is not an argument on the other side of this—nobody is arguing in favor of bias. However, the path to minimizing bias is not always clear.

\section{Fairness in Context}
A machine learning model might reduce bias in one sense, such as in the distribution of predictive scores across racial groups, while not reducing bias in a different sense, such as in the likelihood across racial groups of the outcome occurring. It could even be that reducing bias in one area makes bias worse in another area. The ethical stance of fairness alone does not provide guidance on what to do when fairness in one aspect comes at the expense of fairness in a different aspect. Further, it does not say how much accuracy a machine learning model should be willing to sacrifice in order to achieve greater fairness. How important is fairness for a model that is fundamentally not accurate?
In justice settings in particular, where the best features for predicting future justice system involvement are also all correlated with race, there is no easy path to manifesting the axiom of fairness in actual machine learning practice. What is unbiased machine learning when all available features are biased?
This tension becomes even more crystalline when considering machine learning errors. In particular, the frequency of false positives might be higher than the frequency of false negatives for one group but be reversed for another group. Because the base rate of the predicted outcome (re-arrest) is markedly higher for one group than another group, false positive rates and false negative rates are going to be different for each group. In this case, how should fairness be understood? Especially in justice settings, where false positives mean something very different in terms of costs than false negatives, is fairness achievable when one group has more false positives and another has more false negatives?

\section{Setting False Positive and False Negative Weights}
All machine learning modeling must take some stand on the weights of false positives and false negatives. Often machine learning algorithm packages set equal weights for false positives and false negatives. For example, the \emph{sklearn} package for Python \cite{b3} allows for the setting of "class\_weight" which has a default weight of 1 for all classes. Otherwise specific weights can be applied. The setting also allows for a "balanced" mode that sets weights inversely proportional to class frequencies. Similarly, the caret package for R \cite{b4} allows weights to be set to impose a heavier cost when errors are made in the minority class.
While weights can be a parameter for training a machine learning model, they are not computationally solved. The setting of weights on false positives and false negatives is a normative decision that the data scientist must make. The algorithm allows for different weights, but there is not computationally determined "best" weights. The decision on weights, including the decision to leave the default in place, is based on some valuation of which is worst: is it worse for the model to produce false positives or false negatives. Analytics tools, like the Receiver Operating Characteristic, make this point clear. The ROC curve shows how trade-offs might be made, where the boundary of possible trade-offs lies, but it does not decide which point on that curve best fulfills the values and mission of the organization or the project.

\section{Equal Weights?}
In some cases, false negatives and false positives might warrant equal weights. Because both represent errors, they both signify something about the algorithm's accuracy. A determination to maximize accuracy might not care about a distinction between false negatives and false positives as they are both errors detracting from accuracy. A modeling effort that improved its AUC score might not care whether that improvement came from reducing false positives or false negatives.
In some settings, though, like medical diagnostics for example, the difference between false negatives and false positives might play an important role. Whether a positive diagnostic test should result in a new medical procedure, for example, might depend greatly on what kind of error the diagnostic test produces and the potential upsides and downside of the medical procedure itself. Also, weighing false positives and false negatives might depend on how invasive the procedure was and what the likelihood of success was. There are no hard and fast rules for deciding how much weight to put on false positives and false negatives. There is no computational solution. Weights have to be decided on the basis of values and mission, not on data and calculation.

\section{Justice Settings}
In justice environments, decision support tools developed from statistical modeling and machine learning have become the professional standard. These tools can inform justice decisions around detention, pre-trial release, dispositions and sentencing, probation, parole, safety and security inside an institution, release, among others. While there are some common features (like legal history), they vary by content, by length, by calculation mechanism, among other things. In all cases, though, they make recommendations about either further penetration into the justice system, or a path toward justice system exit, or simply to release someone without any further intervention.  These tools sometimes work as a triage mechanism, sorting people into different levels of intervention. At other times, they work to inform degree of consequence of punitiveness. Often they are established as a direct result of machine learning, other times statistical findings are moderated through local preferences and stakeholder values.
There are two questions about false positives and false negatives that are especially critical for those creating machine learning models in justice settings. One, in a justice setting, are false positive and false negatives equally significant? And two, in a justice setting, who ought to determine the trade-off between more or less false positives of false negatives?
In the justice setting, the difference between a false positive and a false negative is significant. A false positive can mean over-intervention. It can mean an unnecessary and harmful application of punitive measures, like detention or placement in a corrections facility. In this sense, a false positive is largely negative. Interventions can have negative consequences for individuals, families, and communities. The repercussions of interventions like probation or placement can make educational achievement much less likely, can make employment less secure \cite{b5}, and can make future justice system involvement more likely. One study found that in the year following incarceration, only 55 percent of people had any income at all \cite{b6}. One study found that any contact with the juvenile justice system can lead to double the likelihood of being arrested as an adult, compared to youth with similar behaviors who were not involved with the justice system \cite{b7}. And incarceration can have impacts across the life span, including on divorce and separation \cite{b8} and on things like worse health outcomes, financial struggles, and impaired social standing \cite{b9}. In this way, false positives are clearly negative and are not insignificant.
Justice system intervention can also include supportive interventions. Interventions like mentorship, community based positive supports, or employment readiness have the potential to be helpful, even for people who are unlikely to be re-arrested. Depending on how welcome and how restrictive these kinds of positive supports are, this kind of false positive is likely to be less harmful than interventions based on control and the restriction of individual liberty. False positives might also be a waste of resources, targeting interventions to someone who was not otherwise going to get into trouble again. For prevention goals, there is nothing to prevent if the person was not likely to be re-arrested in the future.
Reciprocally, false negatives can have a different set of important consequences. False negatives occur when a justice system decides to not intervene for a person who actually was likely to later be re-arrested or referred into the justice system. This error is most costly when the justice system does not intervene and the person goes on to hurt someone in the community, causes harm in the community, or is arrested for a new law violation. The impact of a false negative has the potential to be large. People working in justice systems can sometimes find these kinds of events to be the hardest part of the work. It is understandable that no one would want their well-intentioned decision to lead to harm in their community.
Both false positives and false negatives have the potential to be costly to individuals and to the community. Part of the difficulty of assigning weights to them is that false positives (over-intervention) tend to occur much more often than the kind of false negative (under-intervention) that really stirs a community. Without a decision support system, a risk assessment, or a machine learning model, there is sometimes an incentive to over-intervene; it may be better to accept many smaller costs borne by separate individuals than to feel the consequences of one large cost borne by the whole community.

\section{Do Justice Systems Pay Attention to False Positives and False Negatives?}
Regardless of how the details of different costs associated with over- and under-intervention might be tabulated, most justice systems do not explicitly take a stance on which kind of error is worse than the other, and by how much. Not only does this kind of calculation not occur broadly, it does not occur in a targeted way that would lead to specific class weights for justice system decision support systems, risk assessments, and machine learning models. Scanning risk assessment validation reports, it is apparent that the focus is most typically on overall accuracy as indicated by the area under the Receiver Operating Characteristic curve, or AUC, or maybe simple correlations. For example, the "Validation Study of the Youth Level of Service (YLS) Assessment in Hennepin County" \cite{b10} does not examine errors at all and does not mention false positives or false negatives. Similarly, the "Long-Term Validation of the Youth Assessment and Screening Instrument (YASI) in New York State Juvenile Probation" \cite{b11} reports area under the curve statistics, but does not examine false positive or false negative rates either overall or across different version of the tool. Little consideration is given to whether weights were determined for false positives and false negatives, or how weights were determined, or how different weights might impact accuracy.

\section{Base Rates and Race}
Base rates are the rate at which an outcome occurs for someone who is currently in the justice system. It is a prospective likelihood for someone who has already been referred, or arrested, or adjudicated, or tried, or sentenced, about whether they will return to the justice system in the future. It is sometimes defined around re-referrals, re-arrests, re-adjudications, etc. Base rates are commonly different across racial groups. For Black people and for White people, in particular, there is a persistent difference across time and place in base rates.
Base rates for people who are in the justice system currently are different from more general outcome rates or even recidivism rates. Base rates are only for a recurrence, not an initial occurrence. They are not for the overall population, just for the system involved group. And they are not for the lifetime of the person, usually only one or two years. They are measured by following a cohort of people who entered the justice system during one time and then following them to check whether they re-enter the justice system during some follow-up period of time. There are no set standards for risk assessment follow up time periods, which outcomes to track, or which types of events relate to the index event versus are independent events. Because of this, it is difficult to compare base rates between models, jurisdictions, or over time.
Lower base rates might lead to more false positives. In justice settings, it is the norm for one racial group (such as White youth) to have lower base rates of the outcome occurring than others (such as Black youth). This difference in base rates between groups is likely not due to race. Differences by race are likely due to structural factors that differ between the racial groups.
The crucial importance of base rates in justice setting models is the difference by race. Many model studies do not report base rates by race at all (see " Evaluating the predictive validity of the COMPAS Risk and Needs Assessment System" \cite{b12}, "The Criminal Court Assessment Tool: Development and Validation" \cite{b13}, "Re-validating the Federal Pretrial Risk Assessment Instrument (PTRA): A Research Summary" \cite{b14}, or "Long-Term Validation of the Youth Assessment and Screening Instrument (YASI) in New York State Juvenile Probation" \cite{b15} as just a few examples), but some do. One study found that the Florida PACT showed a re-adjudication rate of 40.4 percent for Black youth and 33.8 percent for White youth \cite{b16}. The same study found that the Virginia YASI showed a base rate of 19.9 percent for White youth and 33.0 percent for Black youth. The "Validation of Virginia's Juvenile Risk Assessment Instrument" \cite{b17} reported that the 12-month re-arrest rate was 41.1 percent for Black youth and 28.4 for Black youth. In sum, many risk assessment validation studies do not examine base rates by racial group at all, and those that do look at base rates by racial group report markedly higher base rates for Black people than for White people.

\section{Base Rates and Error Types}
The importance of examining base rates by race is that differences in base rates translate directly into differences in false positives and false negatives. If a model predicts that 40 percent of all who are classified as high-risk are likely to experience a re-arrest, and if White people experience the outcome at a rate of 30 percent while Black people experience the outcome at a rate of 50 percent, then there will be more false positives for White people and more false negatives for Black people. It is not a failure of the machine learning algorithm that produces different frequencies of false positives and false negatives for different racial groups. It is not a computational error. The reason for the different error rates is that the base rates themselves are different.
The problem of different rates of false positives and false negatives across racial groups in justice settings is twofold. One, base rates differ across racial groups because of historical and structural racism. Structural racism from things like policies of redlining neighborhoods \cite{b18}, of branch banking \cite{b19}, of policing \cite{b20} of broken windows and stop and frisk \cite{b21}, among others, can lead to higher rates of arrest for Black people than for White people. The relationship between crime rates and race is complex, and a review of that extensive literature is beyond the scope here. However, differences in base rates across racial groups are not free from an underlying prejudice bias in the data. The systems that produce crime data are not free from institutional and structural bias. Two, the justice system in the United States is overwhelmingly disproportionately Black. In a system that has become more, not less, racially disparate in recent decades, what does it mean for decision support tools to work differently for Black people and for White people? Does the cost of a false positive or of a false negative mean something different for Black people than for White people? Does a legacy of over-involvement of Black people in criminal justice count in the arithmetic of decision-making errors?

\section{Achieving Fairness}
Fairness in machine learning in the justice system cannot be achieved through computation alone. Better machine learning algorithms, more features, or rebalanced samples will not be enough to claim fairness. And more sophisticated statistical methods, like latent variable analysis, structural equation modeling, propensity score matching, or factor analysis will also not be enough to sate the demand for fairness.
The only path for machine learning to be fair, is to consider how fair the underlying decision making is. Careful consideration of each decision point, with input from leaders, workers, stakeholders, and people from impacted communities, can help clarify what the purpose of the system is, what the purpose of an intervention is, what the goal of punishment is. Government agencies can move fairness conversations forward, with support from the data and from machine learning models, only by having courageous conversations with a broad set of voices about what a community wants and what it is willing to tolerate to get what it wants.


\begin{thebibliography}{00}
\bibitem{b1} S. Verma, and J. Rubin, "Fairness definitions explained," IEEE/ACM International Workshop on Software Fairness (FairWare), 2018, pp. 1-7.
\bibitem{b2} A. Cutler, M. Pribić, and L. Humphrey, "Everyday ethics for artificial intelligence," IBM Corporation, 2019.
\bibitem{b3} F. Pedregosa, G. Varoquaux, A. Gramfort, V. Michel, B. Thirion, O. Grisel, M. Blondel, P. Prettenhofer, R. Weiss, V. Dubourg, and J. Vanderplas, "Scikit-learn: Machine learning in Python", the Journal of machine Learning research, 2011, pp. 2825-2830.
\bibitem{b4} M. Kuhn, J. Wing, S. Weston, A. Williams, C. Keefer, A. Engelhardt, T. Cooper, Z. Mayer, B. Kenkel, R.C. Team, and M. Benesty, "Package ‘caret’," The R Journal, 2020.
\bibitem{b5} A. Looney, and N. Turner, Work and opportunity before and after incarceration, Washington, DC. Brookings Institution, 2018.
\bibitem{b6} C. Wildeman, and B. "Western, Incarceration in Fragile Families", The future of children, 2010, pp. 157-177.
\bibitem{b7} A. Aizer, and J.J. Doyle Jr, "Juvenile incarceration, human capital, and future crime: Evidence from randomly assigned judges," The Quarterly Journal of Economics, 2015, pp. 759-803.
\bibitem{b8} M. Massoglia, B. Remster, and R.D. King. "Stigma or separation? Understanding the incarceration-divorce relationship," Social Forces, 2011, pp. 133-55.
\bibitem{b9} L.C. Porter, "Incarceration and post-release health behavior," Journal of Health and Social Behavior, 2014, pp. 234-249.
\bibitem{b10} Hennepin County, Validation Study of the Youth Level of Service (YLS) Assessment in Hennepin County.
\bibitem{b11} Long-term Validation of the Youth Assessment and Screening Instrument (YASI) in New York State Juvenile Probation, Orbis Partners, Inc., 2007.
\bibitem{b12} T. Brennan, W. Dieterich, and B. Ehret. "Evaluating the predictive validity of the COMPAS risk and needs assessment system," Criminal Justice and Behavior, 2009, pp. 21-40.
\bibitem{b13} S. Picard-Fritsche, M. Rempel, A. Kerodal, and J. Adler. The Criminal Court Assessment Tool: Development and Validation, New York, NY, Center for Court Innovation 2017.
\bibitem{b14} T.H. Cohen, C.T. Lowenkamp, and W.E. Hicks, "Revalidating the Federal Pretrial Risk Assessment Instrument (PTRA): A Research Summary," Federal Probation Journal, 2018, pp. 23-39.
\bibitem{b15} Long-term Validation of the Youth Assessment and Screening Instrument (YASI) in New York State Juvenile Probation, Orbis Partners, Inc., 2007.
\bibitem{b16} C. Baird, T. Healy, K. Johnson, A. Bogie, E. Dankert, and C. Scharenbroch, A Comparison of Risk Assessment Instruments in Juvenile Justice, Madison, WI, National Council on Crime and Delinquency, 2013.
\bibitem{b17} J.P. Schneider, Validation of Virginia's Juvenile Risk Assessment Instrument, dissertation, Virginia Commonwealth University.
\bibitem{b18} A.E. Hillier, "Redlining and the Home Owners' Loan Corporation," Journal of Urban History, 2003, pp. 394-420.
\bibitem{b19} T. Morris, "Branch Banking and Institutional Racism in the US Banking Industry," Humanity and Society, 2008, pp. 144-167.
\bibitem{b20} A. Mesic, L. Franklin, A. Cansever, F. Potter, A. Sharma, A. Knopov, and M. Siegel, "The Relationship Between Structural Racism and Black-White Disparities in Fatal Police Shootings at the State Level," Journal of the National Medical Association, 2018, pp. 106-116.
\bibitem{b21} J. Oberman, K. Johnson, "The Never Ending Tale: Racism and Inequality in the Era of Broken Windows," Cardozo Law Review, 2015, pp. 1075–1091.

\end{thebibliography}
\end{document}